\documentstyle[aps,prb,twocolumn]{revtex} 

\begin{document}
\draft
\title{Reduced O diffusion through Be doped Pt electrodes}
\author{R. Stumpf$^*$, C.-L. Liu, and C. Tracy} 
\address{Motorola, Embedded Systems Technology Laboratories, 
Los Alamos, NM\,87545-1663}

\date{\today}
\maketitle

\begin{abstract}
Using first principles electronic structure
calculations we screen nine elements for their potential to retard
oxygen diffusion through poly-crystalline Pt (p-Pt) films.  We determine
that O diffuses preferentially as interstitial along Pt grain boundaries
(GBs). The calculated barriers are compatible with experimental
estimates. We find that Be controls O diffusion through p-Pt.  Beryllium
segregates to Pt GBs at interstitial (i) and substitutional (s) sites.
i-Be is slightly less mobile than O and it repels O, thus stuffing
the GB. s-Be has a high diffusion barrier and it forms
strong bonds to O, trapping O in the GB. Experiments confirm our
theoretical predictions.
\end{abstract}
\pacs{64.75.+g,66.30.-h,61.72.Mm}

A serious problem preventing the use of high dielectric oxide materials
(e.\,g.\ BaSrTiO$_3$) for capacitors as part of dynamic random access
memory devices is the O diffusion through the electrodes and
subsequent oxidation under the electrodes. Pt electrodes do not oxidize
but they allow O to diffuse through the p-Pt film during deposition and
anneal of the dielectric.  This causes unwanted oxidation below the Pt
film.  It has been proposed (but not yet demonstrated) that O diffusion
can be reduced by doping of the Pt film.\cite{Grill95} The addition of
barrier layers or the use of different electrode materials are other
attempts to cope with the oxidation issue.\cite{Kotecki97}

All the theoretical results reported in this study are obtained using
the first principles total-energy code VASP (Vienna {\em Ab-initio\/}
Simulation Package)\cite{vasp} within the generalized gradient
approximation (GGA).\cite{Perdew92} Electronic wavefunctions are
expanded in a plane wave basis. The atomic cores are represented by
ultrasoft pseudopotentials\cite{Vanderbilt90} which allow for a reduced
plane wave basis set (cut-off 270\,eV).

It has long been assumed that O diffuses
along grain boundaries in the p-Pt films used as
electrodes.\cite{Schmiedl96} The model GB we choose for our diffusion
study is the $\Sigma5(310)[001]$ symmetric tilt GB (See Fig.\,1). To
determine the O diffusion mechanism from theory we first have to find
the preferential site for O in Pt (see Table~1). The calculated binding
energy for the O$_2$ molecule is 4.91\,eV per O atom.\cite{O2molecule}
The most stable site for O in bulk fcc-Pt is the tetrahedral
interstitial with a binding energy of 3.45\,eV.  Interstitial sites at
the $\Sigma5$ GB have binding energies up to 4.60\,eV. Thus O in Pt
strongly segregates to GBs. In agreement with experiment our results
indicate that Pt does not oxidize and even the O in GBs is unstable with
respect to the formation of gaseous O$_2$.\cite{O2molecule}

Any diffusion of O involving sites in bulk Pt has a high activation
energy of at least 2.64\,eV.
To diffuse through bulk-Pt, O first has to move from the GB to a bulk
site which costs 1.15\,eV as bulk-interstitial or 2.17\,eV as
bulk-substitutional. For O interstitial migration one has to add the
tetrahedral-octahedral energy difference of 1.49\,eV. Substitutional O
likely migrates with the help of a Pt vacancy\cite{Adams89} which has a
calculated formation energy of 0.73\,eV. Thus crystalline Pt films
act as O diffusion barriers. If such films could be deposited reliably,
at low stress, and with good adhesion the O inter-diffusion problem would
be solved.

To calculate the diffusion barriers for interstitial O in the GB we use
two different techniques. In the traditional approach we determine the
potential energy surface (PES) for an O atom moving within the GB. This
is done by calculating the total energy of the GB system with an O
interstitial fixed at the points of a rectangular 10$\times$6 grid
spanning the irreducible (310) interface cell. The O's coordinate
perpendicular to the GB plane and the position of most of the Pt atoms
are relaxed at each mesh point. Four Pt atoms distant to the O are fixed
to prevent a rigid translation of the Pt film.  The resulting potential
energy surface (PES), Fig.\,2, indicates one main minimum at coordinate
(5.5, 4.5), at least one secondary minimum at (9,4), and a clear
diffusion path in the [001], i.e. the short, direction with a barrier
(at 5.5,1) of about 0.7\,eV. In the [$\bar1$30] direction the O has to
cross at least two saddle points. The highest saddle is at (8,4.5) and
appears to have a barrier of about 0.9\,eV.

Application of the nudged elastic band (NEB)\cite{Jonsson98} method
leads to a more accurate determination of the O diffusion barriers in
the $\Sigma5$ GB.  In the [001] direction the barrier is 0.68\,eV and in the
[$\bar1$30] direction we identify two saddle points with almost the same
barriers: 0.68\,eV at (1,3) and 0.67\,eV at (8,2). Note that the low
barrier at (8,2) is not obvious from Fig.\,2. Also, the full variability
of the energy along the diffusion path in the [$\bar1$30] direction is
not reflected in the calculated PES. The NEB is more reliable in
predicting diffusion paths and barriers.

It is not clear if diffusion along the chosen $\Sigma5$ GB is
representative of O diffusion in p-Pt. Thus we compare
our results to experimental findings. In the most careful study of O
diffusion through nanocrystalline Pt films Schmiedl et al.\ have
determined the diffusion rate of O at room temperature.\cite{Schmiedl96}
The measured value is about $D=10^{-19}$cm$^2$/s, depending on the
microstructure of the Pt film, especially the density of GBs. This
dependence indicates that GB diffusion is the dominant diffusion
mechanism. To compare with the calculated value we estimate an O
diffusion barrier $E_d$ from the experimental diffusion rate
\begin{equation}
D = D_0 e^{-E_d/k_BT}\;.
\end{equation}
The grain size of the Pt films grown by Schmiedl et al. is about
100\,\AA. This means that only about 1/100 of the area of the film
allows O diffusion. Thus the diffusion rate $D_{GB}$ in the GB is about
10$^{-17}$\,cm$^2$/s. A typical diffusion prefactor $D_0$ is
$10^{-3}$\,cm$^2$/s and $k_BT$ is 0.025\,eV. Solving for $E_d$
results in $E_d = 0.8$\,eV. The error margin for this estimate is at
least 0.2\,eV. We conclude that the calculated diffusion barrier of
0.68\,eV is compatible with experiment and that O diffusion in p-Pt
proceeds along GBs.

The main goal of this study is to identify dopants that reduce the O
diffusion through p-Pt films and thus keep the oxidation of material
below the p-Pt at tolerable levels. Potential dopants have to meet
certain conditions. For example, the dopant element should have a high
melting point so that sputter targets can be produced easily.  In
addition, the dopant should not form a volatile oxide. From the elements
that pass those pre-conditions we choose Be, B, Mg, Ti, V, Cu, Rh, Ta,
and Ir for initial investigation.

Only if a dopant segregates to the GB can it affect O GB-diffusion
effectively.  To check which elements segregate to Pt GBs we calculate
the dopant's binding energy in the stable elemental phase, in a
fictitious Pt$_3$-dopant alloy, at substitutional and interstitial sites
in a Pt bulk matrix, and at different substitutional and interstitial
sites at the Pt GB (see Table\,1).  Only the binding energies for the
most stable sites out of five substitutional and three interstitial
sites at the GB are listed.  We find that only the two smallest 
elements, Be and B segregate strongly to the grain boundaries and we
determine their effect on O diffusion.

A dopant at the GB can reduce O diffusion as interstitial (i) or
substitutional (s) species. If the dopant prefers i-sites and if the
dopant or a stable dopant-O complex diffuses more slowly than the O
alone the dopant blocks the O diffusion by ``stuffing'' of the GB.
Boron goes interstitial and it repels O in the GB. The calculated i-B
diffusion barrier in the GB is 0.54\,eV, thus it diffuses faster than
O. Most likely O exposure of B doped p-Pt films causes out-diffusion of
the B from the GB to form B$_2$O$_3$ at the surface. In the B-free films
the O diffusion is not affected. This scenario agrees with
experiment.\cite{Grill95}

If the dopants prefer s-sites at the GB then a slow diffusion of the
dopant is almost guaranteed. Diffusion of substitutional species in
close packed materials typically involve vacancies.\cite{Adams89}
Calculated vacancy formation energies at the Pt $\Sigma5$ GB range from
0.72\,eV to 1.08\,eV, i.\,e.\ they are at or above the calculated
vacancy formation energy in bulk Pt. Adding the vacancy migration
barrier of about 1\,eV results in a diffusion activation energy of at
least 1.5\,eV. If the dopant-O interaction is sufficiently different
from the Pt-O interaction in the GB a substitutional dopant reduces O GB
diffusion. A comparatively attractive dopant traps diffusing O atoms
directly, a repulsive dopant traps the O's at sites with no dopant
neighbors.

Beryllium reduces O diffusions as i- and as s-dopant at Pt $\Sigma5$
GBs. Isolated i-Be at the GB experiences an isotropic migration barrier
of 0.70\,eV as determined by a set of NEB calculations.  In the vicinity
of s-Be the barrier increases to about 0.8\,eV caused by a small
repulsive interaction. A neighboring i-Be leads to a larger increase
because of a strong short range repulsion between the two i-Be atoms. To
quantify the barrier increase requires the study of a large number of
paths which has not been done here.  The interaction between i-Be and O
is repulsive. The repulsion energy is about 0.5\,eV at neighboring local
minima and about 0.1\,eV at second nearest neighbor sites depending on
the detailed configuration.  This and the slow diffusion of i-Be 
indicates that i-Be retards O diffusion. To confirm this we test two
possible saddle configurations for O diffusion with i-Be near by. We
find that i-Be increases the barrier for O diffusion by about 0.1\,eV.

The interaction between s-Be and O is strongly attractive which leads to
high O diffusion barriers. We calculate the O binding energy at seven
different sites close to a s-Be at the GB. At the new global minimum the
O is 0.90\,eV lower in energy than at the most stable site at the GB
without Be. The diffusion barrier of O is increased from 0.68\,eV to
0.81\,eV in the [001] direction and to above 2\,eV in the [$\bar1$30]
direction. In the lowest energy configuration containing 2 s-Be atoms,
an O atom is bridging between the two s-Be atoms. The barrier for the O
to leave the bridge site is above 2.5\,eV. We note here that in the
presence of O in the GB it is energetically favorable for i-Be to
convert to s-Be that is bound to O, thus trapping the O.

We conclude that Be reduces O diffusion along Pt GBs in various
ways. The actual O diffusion rate in the presence of Be depends on the
Pt microstructure and the Be concentration and absorption site.
We test our theoretical prediction experimentally by measuring the
WO$_3$ formation beneath a clean and a Be doped p-Pt film under O
anneal. The Pt is sputter deposited at 400$^\circ$\,C to a nominal
thickness of 100\,nm on a substrate consisting of 450\,nm of CVD W on a
TiN adhesion layer on SiO$_2$/Si.  The 1.5\,at.\% Be implant
is performed at 40\,keV and 7$^\circ$ tilt with a dose of
10$^{16}$\,cm$^{-2}$.  For comparison we prepare an identical wafer
without the Be implantation.  Each of the two wafers are pre-annealed at
600$^\circ$\,C in a carefully controlled Ar ambient for 30 minutes prior
to the 5 minutes oxidation heat treatment at 500$^\circ$\,C in open air.
XRD analysis revealed WO$_3$ peaks in both samples.  The WO$_3$\,(110)
peak counts of the Be-implanted sample are 3.1 times smaller than those
of the Be free sample. This result indicates that Be implantation indeed
delays the diffusion of O$_2$ through Pt. We expect that the Be effect
would be even bigger if the Be was co-deposited with the Pt such that
more of the Be has a chance to segregate to the Pt GBs.

In conclusion, we show that Be doping of p-Pt films reduces the O
inter-diffusion. This conclusion is based on our first principles
atomistic model and first experimental results. We expect that the
optimization of the Be doping recipe will increase the Be effect
significantly.  Our study shows how first principles based modeling not
only helps to understand fairly complicated processes like
inter-diffusion in poly-crystalline materials; modeling even shows the
way to modify the diffusion properties. Modeling also helps us determine
rules that govern segregation and diffusion properties, e.\,g.\ the role
of atomic size. This will be subject of a longer paper.\cite{Stumpf99}
This work was made possible by advancements in first principles codes
(e.\,g.\ VASP)\cite{vasp} that allow to calculate large systems
containing ``hard'' elements (e.\,g.\ O and Pt) efficiently and
accurately. The NEB\cite{Jonsson98} enables us to calculate diffusion
paths in complicated cases. Finally, we need parallel computers to
consider the hundreds of configurations necessary to solve ``real
world'' materials science problems.

We acknowledge the use of Sandia National Laboratories computing
resources.

\begin{figure}
\caption{Structure of the (310) GB plane. The in-plane unit cell with its
two symmetrical halves is indicated.}
\end{figure}

\begin{figure}
\caption{PES for O in the $\Sigma5$ GB
corresponding to the lower rectangle in Figure 1. The contour
spacing is 0.2\,eV. The PES is calculated on a 10$\times$6 mesh
in a 6.312\AA$\times$1.996\AA\ cell.}
\end{figure}

\begin{table} 
\begin{tabular}{lcccccccccc} 
             & O    & Be   & B    & Mg   & Ti   & V    & Cu   & Rh   & Ta   & Ir\\ 
element      & 4.91 & 3.74 & 5.76 & 1.52 & 7.74 & 8.9  & 3.77 & 7.17 &11.77 & 8.71\\
Pt$_3$X      & 1.92 & 4.72 & 4.79 & 3.93 &10.85 &10.39 & 4.25 & 7.18 &13.89 & 8.40\\
s-bulk       & 2.43 & 4.91 & 4.58 & 4.04 &11.67 &11.31 & 4.23 & 7.15 &15.45 & 8.47\\
i-bulk$_{\rm oct}$ & 1.96 & 3.98 & 7.14 & 1.00 & 6.86 & 7.33 & 0.37 & 2.37 & 9.54 & 3.45\\
i-bulk$_{\rm tet}$ & 3.45 & 2.43 & 6.04 &-2.48 & 5.90 & 6.52 &-1.10 & 0.74 & 8.16 & 1.70\\
s-GB         & 3.97 & 5.33 & 7.10 & 3.98 &11.63 &11.26 & 4.07 & 7.15 &15.37 & 8.56\\
i-GB         & 4.60 & 5.69 & 7.83 & 2.96 &10.44 &10.48 & 3.24 & 6.08 &13.72 & 7.40\\
\end{tabular}
\caption{Binding energy of dopants as elements and at different
sites in Pt. The energy reference is the spin averaged free atom
in GGA.  For substitutional absorption we assume that the removed
Pt atom gains the Pt-bulk cohesive energy.  The elemental phase
of oxygen$^{8}$ is O$_2$ 
and the pure bulk phase otherwise.  
}
\end{table}


\begin{references}
\bibitem{O2molecule} The energy of the O dimer is the calculated spin
polarized value. The free O atom reference is the valence energy of the
spin averaged free O atom within GGA.
\end{references}
\end{document}